\documentclass{iau}

\usepackage{amsmath}
\usepackage{graphicx}
\usepackage{multirow}
\usepackage{xspace} 
\usepackage{color, soul} 

\usepackage{xcolor}
\usepackage[normalem]{ulem}

\definecolor{seagreen}{rgb}{0.190, 0.525, 0.361}

\newcommand{\pt}{\texttt{PETAR}\xspace}

\newcommand{\bse}{\texttt{BSE}\xspace}
\newcommand{\be}{\textit{penetration factor}\xspace} 

\newcommand{\Ms}{M$_{\odot}$\xspace}

\def\subinrm#1{\sb{\rm#1}}
    {\catcode`\_=13 \global\let_=\subinrm}
\mathcode`\_="8000
\def\upsubscripts{\catcode`\_=12 }

\upsubscripts

\renewcommand{\eqref}[1]{(equation~\ref{#1})}

\begin{document}

\lefttitle{Sara Rastello}
\righttitle{Star-Black Holes interactions in Young Star Clusters}

\jnlPage{1}{7}
\jnlDoiYr{2021}
\doival{10.1017/xxxxx}

\aopheadtitle{Proceedings IAU Symposium}
\editors{C. Sterken,  J. Hearnshaw \&  D. Valls-Gabaud, eds.}

\title{Star-Black Hole Interactions in Young Star Clusters}

\author{Sara Rastello$^{1,2}$, Giuliano Iorio $^{1,2}$, Mark Gieles$^{1,3}$, Long Wang$^{4,5}$}

\affiliation{$^{1}$Institut de Ciències del Cosmos (ICC), Universitat de Barcelona (UB), Martí i Franquès 1, 08028 Barcelona, Spain}
\affiliation{$^{2}$Departament de Física Quàntica i Astrofísica (FQA), UB, Martí i Franquès 1, 08028 Barcelona, Spain}
\affiliation{$^{3}$ICREA, Pg. Llu\'is Companys 23, 08010 Barcelona, Spain}
\affiliation{$^{4}$School of Physics and Astronomy, Sun Yat-sen University, Daxue Road, Zhuhai, 519082, China}
\affiliation{$^{5}$CSST Science Center for the Greater Bay Area, Zhuhai, 519082, China}

\vspace{-1em}

\begin{abstract}
Close encounters between stars and black holes can trigger micro-tidal disruption events (micro-TDEs) in dense young star clusters (YSCs). Using direct $N$-body simulations with \texttt{PETAR}, we found that most micro-TDEs arise from few-body multiple encounters. The inferred rate is $\sim 350$-$450$~Gpc$^{-3}$~yr$^{-1}$. Micro-TDEs could be detected both by upcoming surveys such as LSST, expected to observe $\sim 10$-$100$ events per year, and by their gravitational-wave (GW) signals peaking in the deci-Hertz band, detectable with future instruments such as LGWA and DECIGO.


\end{abstract}
\vspace{-1em}
\vspace{-1em}
\maketitle
\vspace{-1em}
\section{Introduction}
Dynamical encounters between stars and black holes (BHs) in dense stellar clusters (SCs) such as YSCs, can lead to a wide variety of outcomes. These range from the formation of detached binaries (e.g., the so-called \textit{Gaia BHs}; \citealt{elbadry2023b,panuzzo,rastello2023,marinpina2024,iorio24}) to  disruptive events such as \textit{micro-TDEs}. The latter occur when a star (of mass $m_*$ and radius $r_*$) passes within the tidal radius ($r_{\rm t}$) of a BH ($r_{\rm t} \simeq r_* \left( {m_{\rm BH}}/{m_*} \right)^{1/3}$ ) and is partially or fully disrupted \citep{rees1988,perets2016}. Micro-TDEs can be viewed as the scaled-down counterparts of tidal disruption events (TDEs) involving super-massive BHs, and they may give rise to luminous, multi-wavelength electro magnetic (EM) flares as well as bursts of GW \citep{Komossa2015,Toscani2020,Toscani2022,toscani2025}. Previous analytical and numerical work suggests that micro-TDEs may occur at non-negligible rates, often triggered or enhanced by binary or higher-order dynamical interactions \citep{perets2016,Rastello20,Rastello2018,kremer2021,ryu23a, }. These events could represent an important class of energetic transients, potentially linked to ultra-long gamma-ray bursts and fast blue optical transients \citep{perets2016,kremer2021,kremer_2022,Kremer2023}.

\section{Methods}
We performed direct $N$-body simulation of a sample of YSCs modeled with masses $10^3$-$10^5\,M_\odot$, half mass radius and density based on observed Galactic clusters \citep{krumholz2019}. Stellar and binary populations were sampled from \citet{kroupa2001} and \citet{moe17}, and initialized using \texttt{limepy} \citep{gieles2015} King models ($W_0=7$). We simulated 3600 realizations at both low ($Z=0.0002$) and Solar ($Z=0.02$) metallicity, including primordial and dynamically formed binaries (See \citealt{Rastello25} for more details.)
. Cluster evolution was followed with the code \pt \citep{Wang2020b}, to which we implemented a micro-TDE prescription (removing stars passing within the tidal radius of a compact object, with 10\% of their mass accreted onto the BH, see appendix A in \citealt{Rastello25}), while stellar evolution was treated with the updated \bse\ framework \citep{hurley02,banerjee20}. All clusters were evolved in a Galactic potential \citep{bovy} on circular orbits and integrated for $\sim 1.5$ Gyr.

\begin{figure*}
 	  \centering
\includegraphics[width=0.99\textwidth]{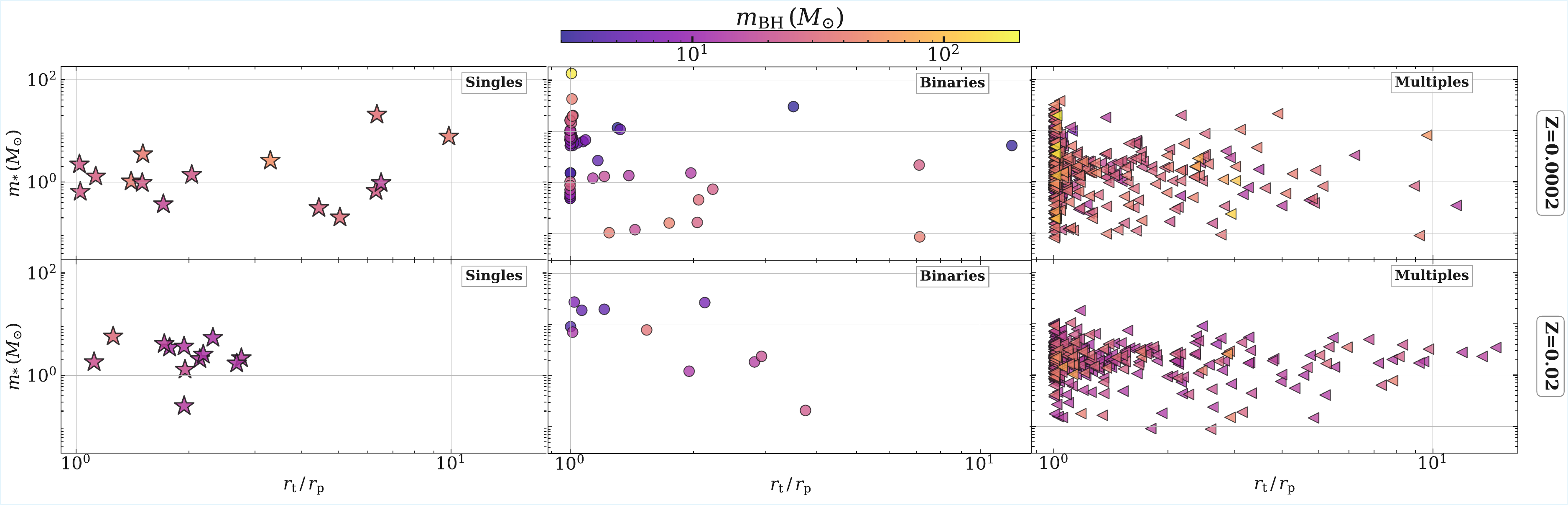}
   \caption{Distribution of \be ($r_{\rm{t}}/r_{\rm{p}}$) and star masses ($m_{*}$ in \Ms) of micro-TDEs occurring during single  (left panels), binary mediated (central panels) and multiple encounters (right panels) for $Z = 0.0002$ (\textit{top panels}) and $Z = 0.02$ (\textit{bottom panels}). The colormap shows the BH masses ($m_{\rm{BH}}$ in \Ms).}
    \label{fig:multipl}
\end{figure*}

\section{Micro-TDEs pathways} \label{sec:mult}
Micro-TDEs in YSCs are produced through three main pathways: i) single encounters, ii) binary-mediated interactions, and iii) few-body multiple encounters. 
Single events are rare ($\sim 3$\%), occurring when a star and a BH approach on a parabolic orbit (Fig.~\ref{fig:multipl}, left panel), typically occurring in dense region (40\% events in within cluster cores). At $Z=0.0002$, BHs are more massive ($17$-$52\,{\rm M}_\odot$), producing deeper encounters ($\beta \sim 1$-$10$), whereas at Solar metallicity they are lighter ($13$-$31\,{\rm M}_\odot$) and less penetrating events occur ($\beta \sim 2$-$3$).
Binary-mediated micro-TDEs (Fig.~\ref{fig:multipl}, central panel) contribute only $\sim 7$\% of the total, mostly ($\sim 82$\%) in metal-poor clusters. They occur either through binary hardening during dynamical encounters or stellar expansion during binary evolution. At $Z=0.0002$, they involve a wide range of BH ($3$-$170\,{\rm M}_\odot$) and stellar (MS and giant) masses ($0.1$-$134\,{\rm M}_\odot$). At Z=0.02, the systems are lighter (BHs: $4$-$35\,{\rm M}_\odot$, stars: $0.2$-$27\,{\rm M}_\odot$). Most binary-mediated micro-TDEs have $\beta \approx 1$ and originate from primordial binaries undergoing common envelope or mass transfer episodes.
The vast majority of micro-TDEs in YSCs ($\sim 90$\%) occur during few-body interactions involving three or more objects (85\% at low-$Z$, 90\% at Z=0.02; (Fig.~\ref{fig:multipl}, right panel). At $Z=0.0002$, disrupted stars are mainly MS ($83$\%), with smaller fractions of giants (8\%), naked He stars (7\%), and WDs (0.5\%); BHs range from $3$ to $138\,{\rm M}_\odot$. At Z=0.02, MS stars dominate even more strongly (98\%), while giants (0.9\%) and WDs (1.1\%) are rare; stellar masses span $0.09$-$18\,{\rm M}_\odot$ and BHs $11$-$75\,{\rm M}_\odot$. Nearly all encounters ($>99$\%) occur on highly eccentric ($e \sim 0.99$) orbits.

\section{Micro-TDEs efficiency \& Rates} 
\label{sec:eta}

\begin{figure}
\centering
\includegraphics[width=0.35\textwidth]{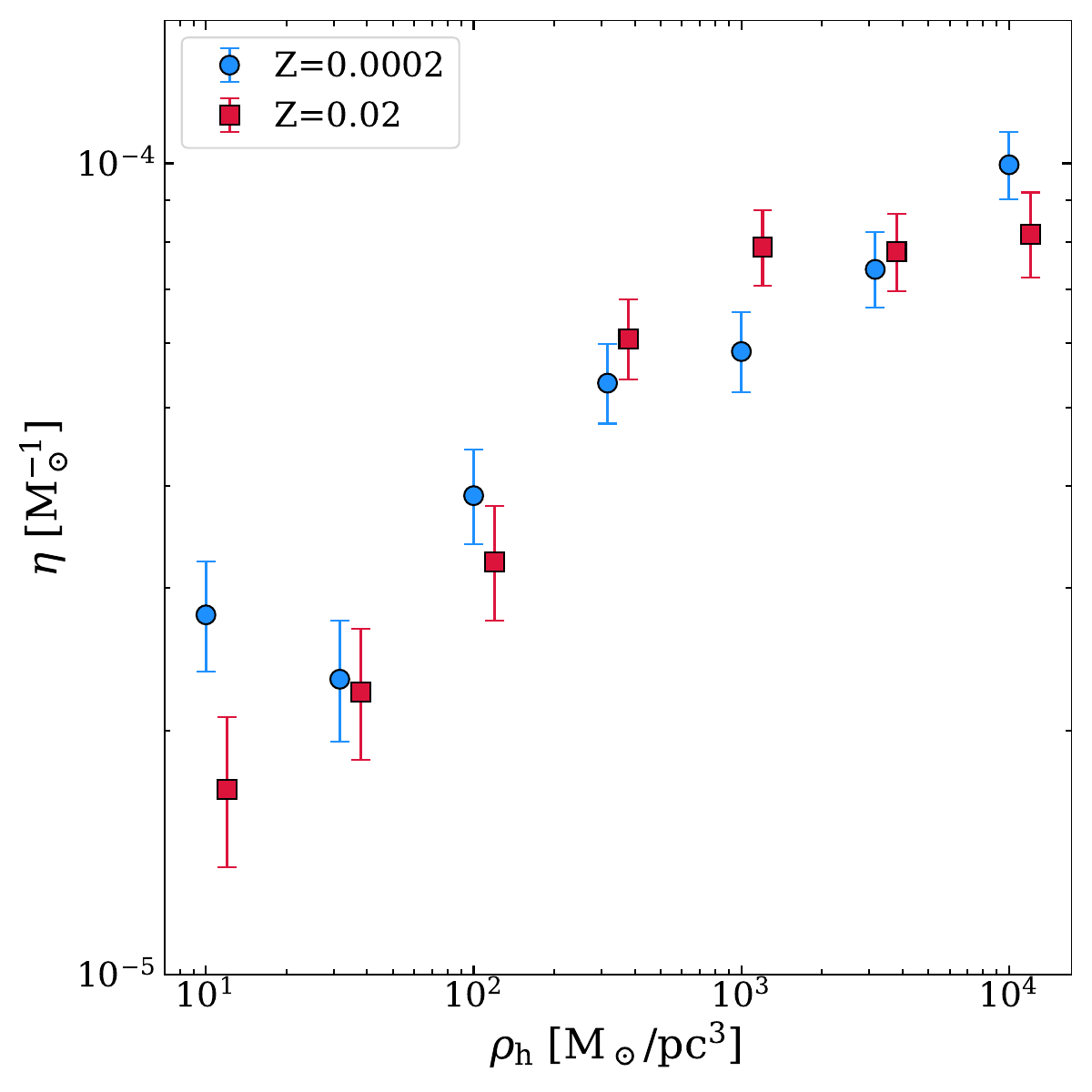}
\includegraphics[width=0.35\textwidth]{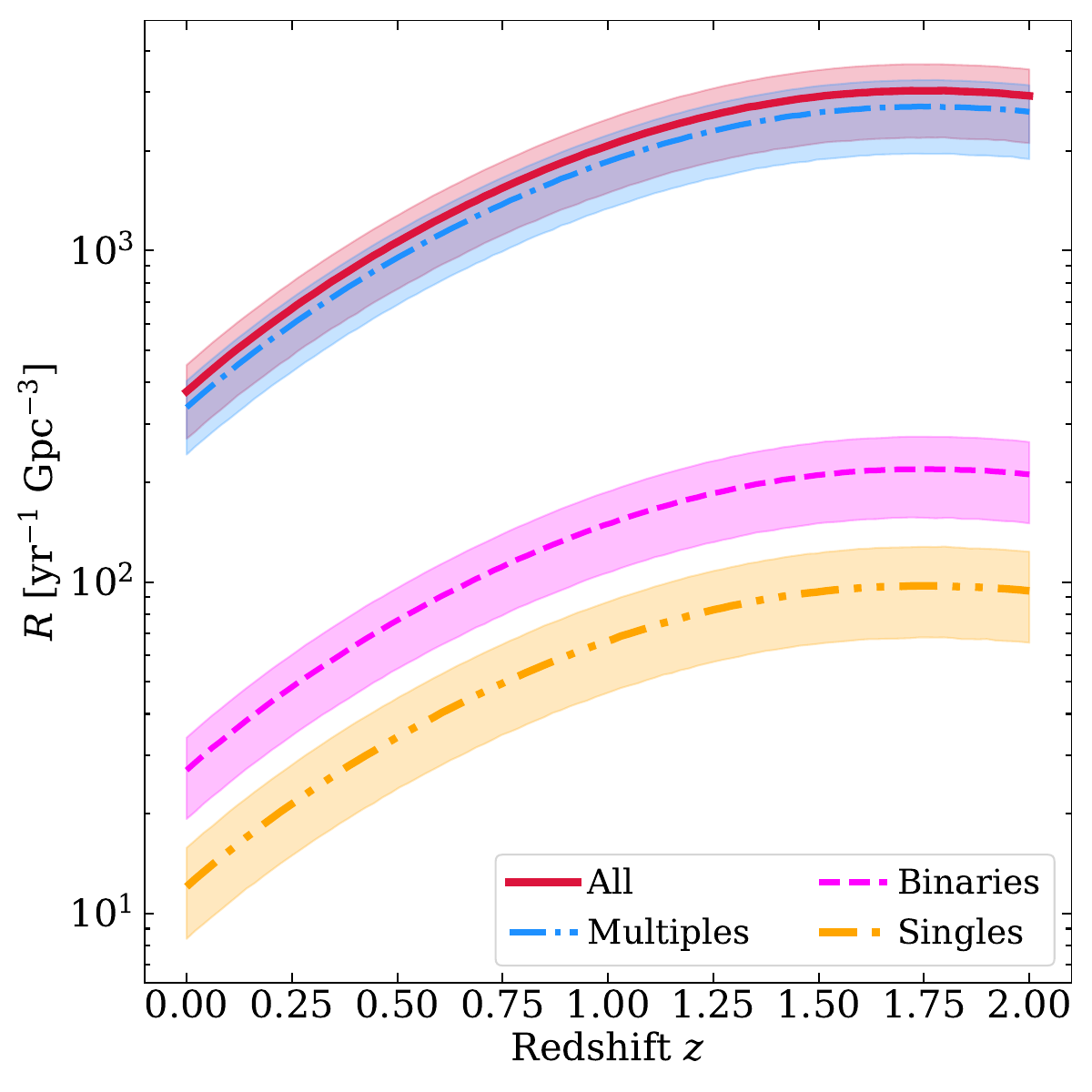}
   \caption{Micro-TDE production efficiencies, $\eta$, as a function of initial cluster density $\rho_{\rm h}$ for $Z = 0.0002$ (blue) and $Z = 0.02$ (red) (\textit{left panel}). Micro-TDE volumetric rates as a function of redshift for each channel (solid lines indicate median rates, while shaded bands the 68\% credible intervals). }
    \label{fig:eff_rho}
\end{figure}

We find a global production efficiency $\eta$ (i.e. the number of micro-TDEs per unit stellar mass) in YSCs of $\eta \approx 5 \times 10^{-5}\,\mathrm{M}_\odot^{-1}$. The distribution (Fig.~\ref{fig:eff_rho}, left panel) shows a strong dependence on cluster initial density but is independent of metallicity. 
From the derived efficiency, we estimated the comoving event rate density (see Eq. 4-6 in \citealt{Rastello25}) as $R(z) = \eta f_{\rm SF,SC}\, \rho_{\rm SFRD}(z)$
where $f_{\rm SF,SC}$ is the fraction of star formation occurring in clusters \citep{Lada2003, kremer2021}, and $\rho_{\rm SFRD}(z)$ is the cosmic star formation rate density \citep{Madau2017}. 
The total volumetric rate of micro-TDEs is therefore estimated to rise from 350-450 Gpc$^{-3}$ yr$^{-1}$ at $z=0$ up to 2000-3000 Gpc$^{-3}$ yr$^{-1}$ at $z=2$, with the majority of events originating from multiple dynamical encounters.
Accordingly, the number of expected events per year rises from just a few within the local Universe ($z < 0.05$, approximately 200~Mpc) to as many as $10^5$ out to redshift $z = 1$.\\

\begin{figure}
 	  \centering
 	  \centering
    \includegraphics[width=0.35\textwidth]{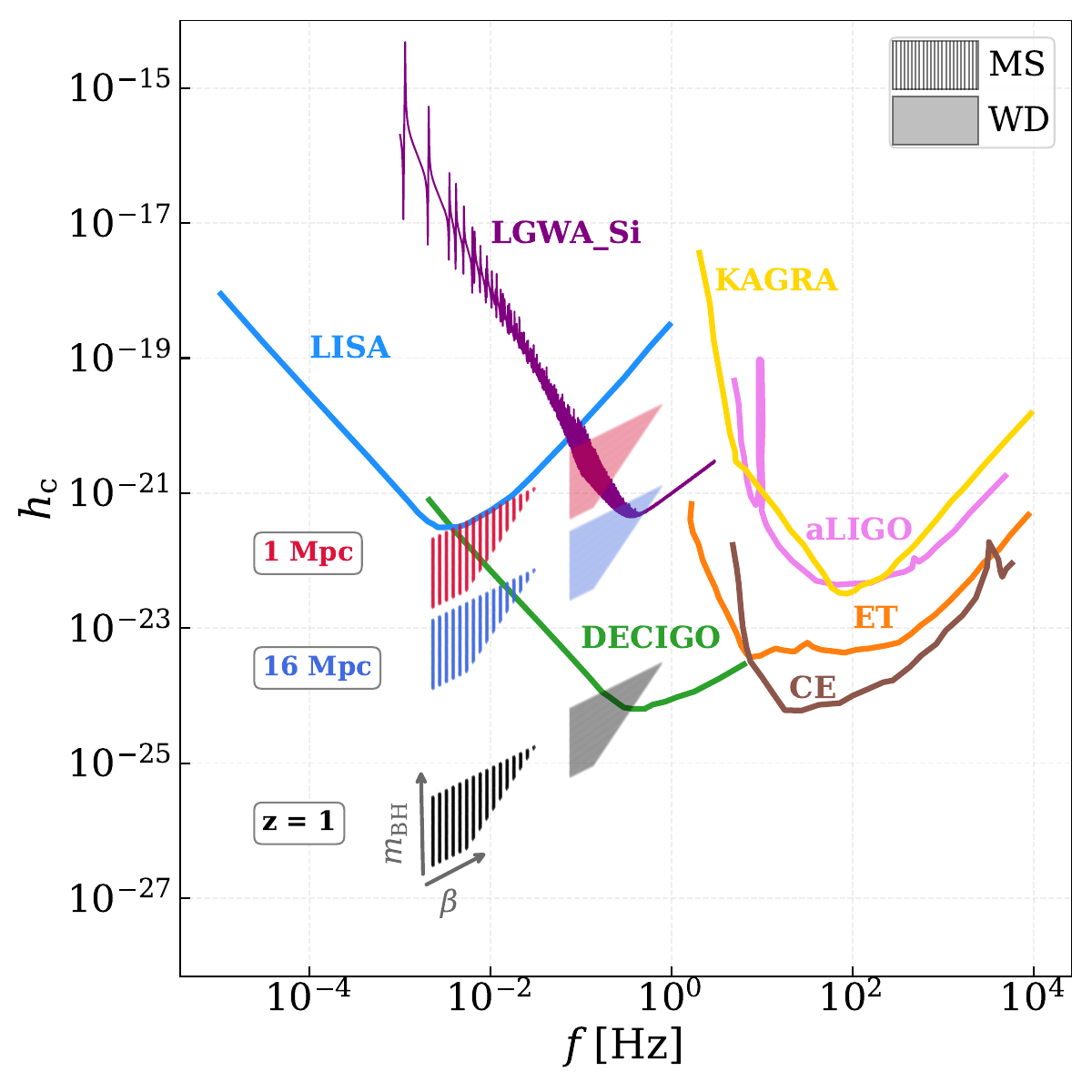}
   \caption{Expected frequency ($f$) and characteristic strain ($h_{c}$) of the GW signal produced by micro-TDEs involving a main sequence star (MS, $m_{\rm MS} = 0.5\,M_\odot$, hatched bands) and a white dwarf (WD, $m_{\rm WD} = 1.4\,M_\odot$, solid colored bands), shown at distance:$\sim$1 Mpc (red), $\sim$16 Mpc (blue), and $z = 1$ (gray). The two gray arrows in the lower-left corner indicate the directions of increasing BH mass ($m_{\rm BH}$) and \be ($\beta \equiv r_t / r_p$). These predictions are overlaid with the sensitivity curves of current and future GW detectors: 
the Lunar Gravitational Wave Antenna (LGWA; \citealt{harms2021}, purple), 
the Laser Interferometer Space Antenna (LISA, \citealt{pau2017}), 
the Deci-Hertz Interferometer Gravitational Wave Observatory (DECIGO; \citealt{sato2017}, green), 
KAGRA (\citealt{abbottGWTC2}, gold), 
the Einstein Telescope (ET; \citealt{maggiore2020}, orange), 
the Advanced Laser Interferometer GW Observatory (LIGO; \citealt{aligo}, violet), 
and the Cosmic Explorer (CE; \citealt{ng2021}, brown). }
 \label{fig:res-gw}
\end{figure}

\section{EM and GW Detectability} \label{sec:detectability}
 
By combining the predicted rates we assessed the detectability of micro-TDEs with current and future surveys as Zwicky Transient Facility (ZTF) at Palomar Observatory \citep{graham2019}, 
Legacy Survey of Space and Time (LSST) at the Vera C. Rubin Observatory \citep{Ivezic2019}, 
and Ultraviolet Transient Astronomy Satellite (ULTRASAT) \citep{Shvartzvald2024} with the wind-reprocessed emission models of \citet{Kremer2023}. These models yield peak luminosities in the range $10^{40}$-$10^{43}\,\mathrm{erg\,s^{-1}}$. 
We find that LSST is the most promising instrument to unveil the micro-TDE population in YSCs, expecting to detect from tens up to several tens of thousands of micro-TDEs per year.\\
Micro-TDEs are multi-messenger sources, and in addition to EM emission, they are expected to produce GW bursts. Following the formalism of \citep{Toscani2020}, we find that WD disruptions by stellar-mass BHs produce GW signals in the deci-Hz band, detectable up to $z \sim 1$ by future detectors such as LGWA and DECIGO (Fig. \ref{fig:res-gw}). MS disruptions yield weaker, lower-frequency bursts, observable only within $\sim 10$ Mpc.      



\section{Conclusion}
Micro-TDEs in YSCs are mostly produced through interactions involving multiple systems, with rates peaking around 
$\sim 10^3\,\mathrm{Gpc^{-3}\,yr^{-1}}$ at cosmic noon, largely independent of metallicity but dependent on cluster density. 
Micro-TDEs are promising multi-messenger transients, potentially detectable by upcoming facilities such as LSST that is expected to detect from tens to several tens of thousands of micro-TDEs
per year, and future deci-Hz GW observatories as LGWA and DECIGO. 

\begin{acknowledgements}
SR acknowledges financial support from the Beatriu de Pinós postdoctoral fellowship program under the Ministry of Research and Universities of the Government of Catalonia (Grant Reference No. 2021 BP 00213). The authors used AI tools (ChatGPT, OpenAI) only for language editing.

\end{acknowledgements}

\bibliographystyle{iaulike} 
\bibliography{bibliography} 

\end{document}